\def\Order#1{{\cal O}(#1)}
\def\mean#1{{\langle{#1}\rangle}}
\def\aka{{\it a.k.a.~}}
\def\ie{{\it i.e.~}}
\def\eg{{\it e.g.~}}
\def\etal{{\it et al.~}}
\def\spose#1{\hbox to 0pt{#1\hss}}
\def\lta{\mathrel{\spose{\lower 3pt\hbox{$\mathchar"218$}}
     \raise 2.0pt\hbox{$\mathchar"13C$}}}
\def\gta{\mathrel{\spose{\lower 3pt\hbox{$\mathchar"218$}}
     \raise 2.0pt\hbox{$\mathchar"13E$}}}
\documentstyle[11pt,aaspp]{article}

\begin{document}

\title{THE PHASE SPACE STRUCTURE NEAR NEPTUNE RESONANCES IN THE KUIPER BELT}
\bigskip
\author{RENU MALHOTRA}
\affil{Lunar and Planetary Institute,
3600 Bay Area Blvd., Houston, TX 77058.\\
{\sl Electronic mail: renu@lpis39.jsc.nasa.gov}}
\bigskip \bigskip
\centerline{To appear in the Astronomical Journal}
\centerline{Submitted: June 1995; revised: August 1995}
\bigskip

%%%%%%%%%%%%%%%%%%%%%%%%%%%%%%%%%%%%%%%%%%%%%%%%%%%%%%%%%%%%%%
\begin{abstract}
The Solar system beyond Neptune is believed to house a population of small
primordial bodies left over from the planet formation process. The region up to
heliocentric distance $\sim 50 AU$ (\aka the Kuiper Belt) may be the source of
the observed short period comets.  In this region, the phase space structure
near orbital resonances with Neptune is of special interest for the long term
stability of orbits.  There is reason to believe that a significant fraction
(perhaps most) of the Kuiper Belt objects reside preferentially in these
resonance locations. This paper describes the dynamics of small objects near
the
major orbital resonances with Neptune.  Estimates of the widths of stable
resonance zones as well as the properties of resonant orbits are obtained from
the circular, planar restricted three-body model. Although this model does not
contain the full complexity of the long term orbital dynamics of Kuiper Belt
objects subject to the full N-body perturbations of all the planets, it does
provide a baseline for the phase space structure and properties of resonant
orbits in the trans-Neptunian Solar system.
\end{abstract}
%%%%%%%%%%%%%%%%%%%%%%%%%%%%%%%%%%%%%%%%%%%%%%%%%%%%%%%%%%%%%%

%\keywords{comets, Kuiper Belt, orbital resonances, solar system dynamics}

\section{Introduction}

It has long been conjectured, on both theoretical and observational grounds,
that the outermost regions of the Solar system may contain a large population
of small icy bodies. For example, on the basis of theoretical considerations of
the genesis of the planetary system from the primordial Solar Nebula,
\cite{Kuiper51} suggested that a remnant population of icy planetesimals left
over from the planet-building process may exist at the current epoch beyond the
orbit of Neptune.  \cite{Whipple64} and \cite{Bailey83} speculated on a
massive comet belt as the source of unexplained perturbations of Neptune's
orbit [although this argument must now be discarded as the post-{\it Voyager}
revisions in the planetary ephemeris no longer show any unexplained residuals
in Neptune's motion (\cite{Standish93})]. \cite{Hamidetal68} analyzed the
orbital plane perturbations of comet P/Halley and concluded that any comet belt
between 40 AU and 50 AU has a total mass less than $1 M_\oplus$. This limit
still allows for very large numbers --- perhaps $\Order{10^9}$ --- of cometary
bodies.

More recently, it has been suggested that the observed short period
comets with orbital periods $\lta 20$ yr --- the ``Jupiter family'' comets ---
originate in a belt of low-inclination bodies just beyond the orbit of Neptune,
between 35 AU and 50 AU [\cite{Fernandez80,FernandezIp83}].
The older hypothesis that short period comets originate in a population of
near-parabolic Oort Cloud comets which are perturbed into shorter orbits by
the giant planets appears unlikely: \cite{Duncanetal88,Quinnetal90} have shown
that the orbital element distribution of the observed short period comets is
inconsistent with a source in the nearly isotropic Oort Cloud but is compatible
with a disk-like source in a trans-Neptune comet belt, which they call the
``Kuiper Belt''.  A possible member of the Kuiper Belt was first discovered in
1992 at a distance of 41 AU from the Sun [1992 QB$_1$, reported in
\cite{JewittLuu93}], and several additional discoveries have been reported
since suggesting a potential population of $\sim35,000$ bodies larger than
$\sim 100$ km [\cite{JewittLuu95}].

The dynamical structure of this population has been the subject of several
recent theoretical studies [\cite{LevisonDuncan93,HolmanWisdom93,Malhotra95}].
The first two of these [\cite{LevisonDuncan93} and \cite{HolmanWisdom93}],
studied the long term stability of test particles in low-eccentricity and
low-inclination orbits beyond Neptune, subject only to the gravitational
perturbations of the giant planets in their present orbital configuration.
They found orbital instability on short timescales ($<10^7$ yr) interior to
33-34 AU, an intricate structure of interspersed regions of stability and
instability in the semimajor axis range of 34 AU to 43 AU, and substantially
stable orbits beyond 43 AU. The intricacy of the dynamical structure appears
to be particularly acute near the locations of orbital resonances with Neptune.

Any primordial trans-Neptune population of planetesimals was undoubtedly
subject to dynamical evolution during the planet formation process, and the
initial conditions assumed in the above studies are not necessarily
representative of the state of the Kuiper Belt at the end of planet formation,
as acknowledged in \cite{HolmanWisdom93}.  The nature of the ``dynamical
sculpting'' of the Kuiper Belt that would have occurred during the late stages
of planet formation was the subject of a study by \cite{Malhotra95}.
This study concludes that the giant planets' orbits would have evolved
significantly and the Kuiper Belt would have been sculpted into a highly
non-uniform distribution early in Solar system history: most of the primordial
small bodies in the region beyond Neptune's orbit and up to approximately
50 AU heliocentric distance would have been swept into narrow regions of
orbital resonances with Neptune, particularly the 3:2 and the 2:1 orbital
resonances which are located at semimajor axes of approximately 39.4 AU and
47.8 AU, respectively.  The orbital inclinations of most of these objects
would remain low (typically $<10^\circ$), but the eccentricities would be
excited to significant values, typically $0.1-0.3$.
This structure would be largely preserved to the present epoch.

Pluto (and its satellite Charon) have long been known to reside in the 3:2
Neptune resonance; the resonance libration protects this pair from close
encounters with Neptune, despite their Neptune-crossing orbit (with
eccentricity
$\sim0.25$).  Several of the newly-discovered Kuiper Belt objects are also
likely residents of this resonance [\cite{Marsden95}].  (However, note that the
orbital parameters of most of these objects remain rather poorly constrained
pending future follow-up observations.)

In any case, it is clear that the locations of Neptune's orbital resonances are
particularly interesting with regard to the long term dynamics of the Kuiper
Belt.

In the region between Neptune's orbit and approximately 50 AU heliocentric
distance, by far the most important orbital perturbations are due to Neptune
alone, although the N-body effects of all the planets -- particularly secular
resonance effects -- also surely play a significant role over the age of the
Solar system. In this paper, I use the simplest possible dynamical model --
namely, the circular, planar restricted three-body model -- to determine the
basic phase space structure near the locations of Neptune's orbital resonances
exterior to Neptune's orbit. This model does not reflect the complexities that
arise with
(i) non-zero orbital inclination of the test particle,
(ii) a realistic representation of Neptune's motion (its time-varying orbit),
and (iii) the perturbations of the other planets.
Some of these unmodeled effects can lead to important variations in the
dynamics. In particular, as mentioned above, secular resonances in certain
regions of phase space can drive the inclinations and eccentricities to large
amplitudes and further complicate the dynamics near mean motion resonances.
Nevertheless, this model is a reasonable first approximation because
(a) the evidence from short period comets suggests a low-inclination source
population,
(b) in the trans-Neptunian space, there are no first order orbital resonances
with any other giant planets, and
(c) Neptune's eccentricity and inclination do not exceed $\sim0.025$ and
$\sim2.5^\circ$, respectively, over billion year timescales under the effects
of the mutual gravitational perturbations of the planets.
The advantage of this simple model is that it is tractable:
it allows for the calculation of 2-dimensional surfaces-of-section in which
the phase space structure --- in particular, the properties of resonance
regions --- can be easily visualized.  This provides for relatively
straightforward estimation of the widths of resonances and the determination
of several interesting dynamical properties of resonant orbits, such as the
libration periods and their dependence on orbital eccentricity and libration
amplitude.  To a considerable degree, most of these properties are preserved
in the more realistic situation.
In short, this model provides a reasonable ``baseline" for the phase space
structure in the trans-Neptune region.

The rest of this paper is organized as follows.
Section 2 describes the technical details of the planar, circular restricted
3-body model for the Sun, Neptune and a Kuiper Belt object, including a
description of the information that may be gleaned from surfaces-of-section.
Section 3 provides a detailed look at the phase space structure near
the major Neptune resonances of interest for the Kuiper Belt.
Section 4 summarizes and discusses the results.

\section{The planar circular restricted three-body model}

The classical planar circular restricted three-body problem is a particular
case of the general gravitational three-body problem of masses
$m_1,m_2,m_3$ defined by the following restrictions:
\begin{enumerate}
\item the motion of all three bodies takes place in a common plane;
\item the third body, $m_3$, has zero mass; therefore it does not influence
the motion of $m_1$ and $m_2$; and
\item the masses $m_1$ and $m_2$ describe circular orbits about their common
center of mass.
\end{enumerate}
In the context of this paper, $m_1$ represents the Sun and $m_2$ represents
Neptune. The system is made non-dimensional by the following choice of units:
\begin{itemize}
\item[--] the unit of mass is taken to be $m_1+m_2$;
\item[--] the unit of length is chosen to be, $a_N$, the constant separation
between $m_1$ and $m_2$ (\ie the mean heliocentric distance of Neptune);
\item[--] the unit of time is chosen such that the orbital period of $m_1$
and $m_2$ about their center of mass is $2\pi$.
\end{itemize}
Then the universal constant of gravitation, $G=1$, and the masses of the Sun
and Neptune are $1-\mu$ and $\mu$, respectively, where $\mu=m_2/(m_1+m_2)$.
For the Sun-Neptune system, we have $\mu=5.146\times10^{-5}$, and this value
is used throughout this paper.

In a reference frame with axes $(X,Y)$ rotating with $m_1$ and $m_2$ and with
origin at their center-of-mass, the Sun and Neptune have fixed coordinates,
$(-\mu,0)$ and $(1-\mu,0)$, respectively, and the third (massless) body has
the following equations of motion:
\begin{eqnarray}
\ddot X &=& 2\dot Y + X -(1-\mu) {X+\mu \over r_1^3}
                        - \mu {X-1+\mu \over r_2^3}, \nonumber\\
& & \label{eq:eqns} \\
\ddot Y &=&-2\dot X + Y -(1-\mu) {Y\over r_1^3}
                        - \mu {Y \over r_2^3}, \nonumber
\end{eqnarray}
where $r_1$ and $r_2$ are the test particle's distance to the Sun and to
Neptune, respectively:
\begin{eqnarray}
r_1 &=& [(X+\mu)^2+Y^2]^{1/2}, \nonumber\\
& & \\
r_2 &=& [(X-1+\mu)^2+Y^2]^{1/2}. \nonumber
\end{eqnarray}
These equations of motion admit a constant of the motion, the
{\it Jacobi integral}, given by
\begin{equation}
C = X^2+Y^2-\dot X^2-\dot Y^2 + 2\bigg( {1-\mu \over r_1}
                                       + {\mu \over r_2} \bigg).
\end{equation}
No other constant of the motion is known.
For future reference, we note that the Jacobi integral can be expressed in
terms of the test particle's semimajor axis, $a$, and eccentricity, $e$:
\begin{equation}
C = {1\over a} + 2\sqrt{a(1-e^2)} + \Order{\mu}.
\label{eq:C}
\end{equation}

\bigskip
\noindent{\it Surfaces-of-section}
\bigskip

The motion of the test particle takes place on a three-dimensional subspace
(defined by a particular value of $C$) embedded in the four-dimensional phase
space, $(X,\dot X, Y,\dot Y)$. The usual two-dimensional surface-of-section
(s-o-s) is defined by
\begin{equation}
Y=0, \qquad \dot Y > 0,
\label{eq:sos}
\end{equation}
and the coordinates on the section are $(X,\dot X)$.
The geometrical interpretation is straightforward: we plot the X coordinate
and velocity of the test particle at every conjunction with Neptune.
For orbits exterior to Neptune (strictly speaking, with mean motion smaller
than Neptune's), this occurs every time the test particle is aligned with the
Sun and Neptune and is on the {\it opposite} side of the Sun from Neptune;
for orbits interior to Neptune (\ie with mean motion larger than Neptune's),
this occurs every time the test particle is aligned with the Sun and Neptune
and is on the {\it same} side of the Sun as Neptune.
For the Solar system beyond Neptune, only the exterior orbits are of interest.

In the s-o-s so defined, {\it periodic} orbits of the test particle appear as
discrete points.  The successive crossings of this surface by a
{\it quasiperiodic} orbit lie on a closed smooth curve, while a {\it chaotic}
orbit fills a two-dimensional area.

As the regions in the neighborhood of orbital resonances are of particular
interest here, the following remarks are in order.

\begin{itemize}
\item
The location of a $(j+k):j$ orbital period resonance is defined by Kepler's
relation between the orbital period and the semimajor axis:
\begin{equation}
a = a_{res} = \bigg({j+k\over j}\bigg)^{2/3}
\label{eq:ares}
\end{equation}
where $j$ and $k$ are integers.
\item
In general, at any resonance, one finds a family of stable periodic
orbits that can be parametrized by the orbital eccentricity (or, equivalently,
the Jacobi integral, if we use Eqn.~\ref{eq:ares} in Eqn.~\ref{eq:C}) such
that the semimajor axis equals $a_{res}$ for all orbits in this family,
and the eccentricity ranges from zero to some maximum value, $e_{max}$;
for $e > e_{max}$ the periodic orbit becomes unstable.
\item
In some neighborhood, $\pm\Delta a$ of $a_{res}$, near each resonance, there
exist stable quasiperiodic orbits that librate with finite amplitude about the
exact resonant orbit; the half-width $\Delta a$ depends upon the mass parameter
$\mu$, the resonance itself (\ie $a_{res}$) and the Jacobi integral $C$.
\end{itemize}

Let us consider the Neptune-Pluto 3:2 resonance as an illustrative example.
If Pluto's inclination is neglected, and if one assumes a fixed circular
orbit for Neptune, then the Jacobi integral $C=2.9798$ for the
Sun+Neptune+Pluto restricted three-body problem.
Figure 1(a) is a s-o-s in the neighborhood of the 3:2 Neptune resonance
for this value of $C$.
This section shows examples of all three types of orbits.
The center of the smooth curves lying near $(X,\dot X)=(-1.65,0)$
corresponds to the periodic orbit whose period is exactly 3/2 that of Neptune.
The smooth curves are quasiperiodic orbits surrounding this exact periodic
orbit and represent test particle orbits that librate about the exact
resonance; the size of a smooth curve bears a direct relationship to the
amplitude of libration. Such librating orbits are phase-protected from having
close encounters with Neptune even if the orbit is Neptune-crossing.
It can be seen that, in general, the regions of quasiperiodic orbits are
surrounded by chaotic orbits, \ie beyond a certain libration amplitude,
the smooth curves dissolve into a chaotic zone. Embedded within the
``chaotic sea'' are other stable zones or chains-of-islands that
represent other nearby higher order and secondary resonances.
The origin of the chaos is due to the overlap of the higher-order and
secondary resonances.

It is perhaps more meaningful to visualize the resonance zone in variables
that can be more directly translated into ``resonance width''. The latter is
traditionally (and somewhat loosely) defined as the range in semimajor axis
where the orbital perturbations are large.  In the restricted three-body
problem with $\mu\ll1$, this is the range in which the amplitude of the orbital
parameter variations of the test particle is not simply linearly proportional
to $\mu$, but is $\Order{\mu^\nu}$ with $\nu\sim0.5$. In general, in the
neighborhood of any resonance, we find regions of stable, librating orbits
surrounded by a substantial chaotic zone where test particles may suffer close
encounters with Neptune. In this paper, I will define the resonance half-width
$\Delta a$ as the maximum amplitude of the semimajor axis variations of
{\it stably librating} resonant orbits.
The value of $\Delta a$ will be estimated from the s-o-s as described below.

The phase space structure seen in the $(X,\dot X)$ plane can also be seen in
other (generalized) canonical coordinates.
For example, in plane polar coordinates the equivalent s-o-s
$(r,\dot r,\theta,\dot\theta)$ in the inertial reference frame is
$(r,\dot r)$ with the same section condition as in Eqn.~\ref{eq:sos}.
The s-o-s condition can be rewritten in terms of polar coordinates as follows:
\[ \theta-\theta'=\left\{ \begin{array}{ll}
                          0^\circ  & \mbox{if $n > 1$}\\
                          180^\circ & \mbox{if $n < 1$}
                          \end{array} \right\}, \qquad\dot\theta < 1   \]
where $\theta'$ is the longitude of Neptune, and $n$ is the mean motion of the
test particle.  In other words, the $(r,\dot r)$ plane is equivalent to the
$(X,\dot X)$ plane with points plotted at every conjunction of the test
particle with Neptune. The $(r,\dot r)$ s-o-s is shown in Figure 1(b).

Carrying this one step further by means of a canonical transformation to
Delaunay variables, $(M,J,\lambda,L$) where $M,\lambda$ are the mean anomaly
and mean longitude and $J=\sqrt{a}(1-\sqrt{1-e^2}), L=\sqrt{a(1-e^2)}$
(cf.~chapter XVII in \cite{BrouwerClemence61} and chapter 10 in
\cite{Goldstein80}), this same section can also be seen in the $(M,J)$ plane.
The $(M,J)$ s-o-s is shown in Figure 1(c).
It is evident from the latter figure that for Pluto-like resonance-locked
orbits (\ie for the particular value of $C$ chosen in this s-o-s) the mean
anomaly at conjunctions with Neptune librates about $180^\circ$ and its
maximum libration amplitude is $\sim\!100^\circ$.  Orbits with larger
amplitudes are chaotic.

In addition to the maximum amplitude of $M$, the resonance libration region
is bounded in $J$ as well. Now, if we note that
(i) the Jacobi integral provides a relationship between $a$ and $e$
(see Eqn.~(\ref{eq:C})), and
(ii) for a particular orbital resonance defined by the ratio $(j+k):j$ of the
orbital periods of the test particle and Neptune, the librating orbits all
have a phase-averaged semimajor axis $\mean{a}=a_{res}$, then any s-o-s can
be labeled uniquely either by the value of $C$ or, in the neighborhood of a
particular orbital resonance, by the value of $\mean{e}$ (with the
understanding that $\mean{a}=a_{res}$).
For example, for the s-o-s in Figure 1, we have $C=2.9798$;
near the 3:2 resonance, $\mean{a}=1.3104$; therefore, from Eqn.~(\ref{eq:C})
we have $\mean{e}=0.250$.  Thus, the phase space structure in the
neighborhood of a particular orbital resonance can be systematically studied
in a set of surfaces-of-section where all of them have a common value of
$\mean{a}$ but various values of $\mean{e}$ (hence different values of $C$).
This allows a visualization of the phase space near an orbital resonance as a
function of the mean orbital eccentricity.

Finally, in Figure 1(d) I have plotted the values of $a$ against $M$ for the
same points represented in Figures 1(a--c).
{\it Note that this is not a surface-of-section.} One might call it a
``pseudo-surface-of-section'', for $a$ is not a canonical variable. However,
the constancy of $C$ imposes a relationship between $a$ and $e$
(cf.~Eqn.~\ref{eq:C}) which ensures that this plot looks very similar to the
$(M,J)$ s-o-s in Figure 1(b) and the stable libration regions and the chaotic
regions are easily distinguished.
The reason for choosing this non-canonical variable is that the resonance
width $\Delta a$ of the stable libration zone can be readily determined from
this figure.  Therefore, the $(a,M)$ plane --- the pseudo-surface-of-section
--- will be used in the next section for a systematic look at the properties
of the phase space in the neighborhood of several Neptune resonances.

The libration timescale in the libration zone and the timescale for orbital
instability in the chaotic zone are of particular interest for the dynamics
of Kuiper Belt objects. These libration timescales have been determined for
individual resonances and are also given in the next section. Typically, the
timescale for orbital instability inside the chaotic zone surrounding a
resonance is only a few times the libration period. However, close to the
boundary between the libration zone and the chaotic zone, this timescale can
become exceedingly large.
In particular, for Pluto-like orbits with eccentricity $\sim0.25$ in the
3:2 Neptune resonance, the timescale for instability in the chaotic zone
(\ie for amplitude of the resonance angle greater than $\sim140^\circ$)
is $\Order{10^5}$ yr, but close to the edge of the libration zone
this timescale increases by an order of magnitude or more. This is
illustrated in Figure 2.

\section{Neptune Resonance Zones}

In principle, there is an infinite number of orbital resonances in the
three-body problem. However, we will see that only a handful warrant
any detailed investigation.

Low-eccentricity orbits in the immediate neighborhood of a planet are unstable.
There is a simple explanation for this orbital instability that follows from
Chirikov's (1979)\nocite{Chirikov79} ``resonance overlap criterion'':
first order $(j+1):j$ resonances with $(j+1) \gta 0.51 \mu^{-2/7}$ overlap
completely [\cite{Wisdom80}].
[Numerical experiments by \cite{Duncanetal89} show that the numerical
coefficient on the right hand side in this relation must be revised to 0.44].
Complete overlap of neighboring resonances results in the destabilization
of all periodic and quasi-periodic orbits in the neighborhood of those
resonances.
In other words, circular test particle orbits are unstable in the immediate
neighborhood $|a-1| \lta 1.5\mu^{2/7}$ of a planet's orbit.
For Neptune ($\mu=5.146\times10^{-5}$), this criterion shows that all first
order resonances with $(j+1)\ge8$ are completely overlapping,
so that circular orbits beyond Neptune with orbital radii less than about
33 AU are unstable.

Therefore, beyond the semimajor axis range of about 33 AU, at most only seven
first order $(j+1):j$ resonances with $j=1,2,...7$ are isolated from each
other (at least for low eccentricities).  However, I have found numerically
that low-eccentricity orbits at the 8:7 and 7:6 Neptune resonances are also
chaotic.  This is not entirely surprising, for the resonance overlap criterion
is quite approximate, and Wisdom's scaling law is strictly valid only in the
asymptotic limit $j \gg 1$.
Therefore, only five first order $(j+1):j$ resonances with $j=1,2,...5$ are of
potential significance for the long-term storage of Kuiper Belt objects.

Furthermore, since all second and higher order resonances in the first-order
resonance overlap region are also destroyed, it follows that there are only
seven second order Neptune resonances of potential interest:
$(j+2):j$ with $j=1,3,5,...,13$. Of these, I have found numerically that
only three (3:1, 5:3, and 7:5) have resonance widths $\Delta a\gta 0.005$.
Therefore, the phase space in the neighborhood of only these three second
order resonances will be discussed here.

Figures 3--10 show the properties of these first and second order Neptune
resonances.  They are arranged in order of decreasing mean heliocentric
distance. Each figure has seven parts: the first four [(a)--(d)] show the
resonance zones in the pseudo-surfaces-of-section [plots of $(a,M)$] for
$\mean{e}=0.1,0.2,0.3$ and 0.4, respectively; note that the vertical scale
is the same in all these figures, so that the variations in the extent of the
libration and chaotic zones with $\mean{e}$ and with the mean distance of the
resonance from Neptune can be readily discerned. The next two panels
[(e) and (f)] show examples of librating, resonance-locked orbits in the
rotating reference frame. The last one, panel (g), shows the period of
libration as a function of the mean orbital eccentricity and the amplitude of
libration of the resonance angle, $\phi$.
A few words are in order regarding the latter variable.

For a $(j+k):j$ resonance
the resonance
angle is defined as follows:
\begin{equation}
\phi=(j+k)\lambda-j\lambda'-k\varpi
\end{equation}
where $\lambda$ and $\varpi$ are the mean longitude and longitude of perihelion
of the test particle, and $\lambda'$ is the mean longitude of Neptune.
(Note that $j$ is a positive integer; $k$ is a negative integer for interior
resonances, positive for exterior resonances; the latter are the only
resonances
of interest in this paper.) $\phi$ is the natural variable that arises in the
perturbative analysis of orbital resonances, and it can be used to make a first
approximation theory that models the resonant motion of the test particle with
a
single degree-of-freedom pendulum-like dynamical system
[see \eg \cite{Malhotra94}].
For a test particle locked in a stable exterior resonance, $\phi$ librates
about a mean value which is usually $180^\circ$.
[There are two exceptions to this:
(i) for sufficiently high eccentricity, librations of the resonance
angle about $0^\circ$ are also possible; and
(ii) the 2:1 and the 3:1 resonances allow for asymmetric librations where
the center of libration of the resonance angle is displaced away from
$180^\circ$.  These are discussed further below.]
At conjunctions with Neptune, the resonance angle is related to the mean
anomaly: $\phi = jM + \Order{e}$. Therefore, the librations or chaotic
variations of $M$ evident in the surfaces-of-section are also reflected
in the behavior of $\phi$.

The physical significance of the resonance angle can be seen by noting that
when the test particle is at perihelion, $\lambda=\varpi$, so that
\[ \phi=j(\varpi-\lambda')\qquad\mbox{(at perihelion).}\]
Thus, the behavior of $\phi$ gives a direct measure of the longitude
separation of Neptune from the test particle's perihelion.
This is obviously a critical quantity, especially for high-eccentricity
Neptune-crossing orbits, as the stable libration of $\phi$ then ensures that
the particle is protected from catastrophic close encounters with Neptune.
The maximum excursion from the center of libration of $\phi$ then gives a
measure of the smallest possible separation between Neptune and the test
particle's perihelion in stable resonance-protected orbits.

The resonance libration period as a function of the amplitude of libration of
$\phi$ is shown for various values of $\mean{e}$ in the last panel (g) of
Figures 3--10.  These curves are rather counter-intuitive.
Recall that in the oft-used pendulum model for non-linear resonances, the
pendulum libration period increases monotonically with amplitude and becomes
arbitrarily large close to the separatrix (\ie the orbit that separates the
oscillations from rotations of the pendulum). In contrast, in Figures 3--10(g),
we see that, except for the asymmetric librations at the 2:1 and 3:1
resonances, the libration period in general {\it decreases} with increasing
libration amplitude.  Furthermore, near a $k$-th order resonance, the leading
order (in eccentricity) resonance term in the perturbing potential for the test
particle is proportional to the $k$-th power of the eccentricity, hence the
small-amplitude libration period might be expected to vary as $\sim e^{-k/2}$;
this analogy from the pendulum model does not hold either, as evident in
Figures 3--10(g). Thus, the standard non-linear pendulum model should be used
with caution for orbital resonances.

The general characteristics of the phase space near these first and second
order Neptune resonances are summarized as follows.
(Note that the 2:1 and the 3:1 resonances are exceptional in many respects,
and are discussed later in detail.)

\begin{itemize}

\item[--] [cf. Figures 3--10 (a--d)]
In general, the phase space in the neighborhood of a resonance contains a
zone of stably librating, resonance-locked orbits surrounded by a zone
of chaotic orbits.  The width of the resonance zone (including the stable as
well as the chaotic regions) increases with $\mean{e}$; however, the width of
the libration zone, in general, shrinks as the resonance separatrix dissolves
into a chaotic layer that increases in thickness with $\mean{e}$.  The typical
width of a Neptune resonance libration zone is $2\Delta a\approx 0.02$, or
about $0.6$ AU.

\item[--] [cf. Figures 3--10 (e)]
The periodic orbit at exact $(j+k):j$ resonance has a $j$-fold symmetry in the
rotating frame: it makes $j$ perihelion passages during a synodic period. The
perihelion (and aphelion) longitudes are spaced $360^\circ/j$.  The
quasi-periodic orbits librating about the exact resonance also exhibit $j$-fold
symmetry in the rotating frame when traced over a complete libration period.
For the stable librators (\ie resonance angle $\phi$ librating about
$180^\circ$), the opposition with Neptune occurs near the test particle's
aphelion; hence these may be called ``aphelion librators''.

\item[--] [cf. Figures 6,7,9,10 (f)]
For sufficiently large $\mean{e}$, a new type of periodic (resonant) orbit is
possible, where one of the $j$ perihelion passages during a synodic period
occurs near Neptune. In fact, the perihelion loop encompasses Neptune. For this
orbit and for the quasi-periodic orbits in its neighborhood, the resonance
angle $\phi$ librates about $0^\circ$. These orbits may be called ``perihelion
librators''.  However, these perihelion librators are probably not of practical
importance, as unmodeled perturbations are likely to destabilize them.

\item[--] [cf. Figures 3--10 (g)]
In general, the libration periods are smaller for resonances located closer to
Neptune, and the libration periods decrease with libration amplitude.  Typical
small amplitude libration periods are (1-2)$\times10^4$ yr.

\end{itemize}

The exceptional cases of the 2:1 and the 3:1 resonances are discussed below.

\bigskip
\noindent{\it The 2:1 Neptune resonance}

The synodic period of a test particle near the 2:1 resonance is approximately
equal to its orbital period.  Thus, conjunctions with Neptune occur every
$\sim330$ years. The properties of the phase space near the 2:1 Neptune
resonance are summarized as follows (cf.~Figure 4).

\begin{itemize}

\item[--]
The resonance half-width, $\Delta a$, increases --- from $\sim 0.001$ to
$\sim 0.018$ --- as $\mean{e}$ increases from $\sim0$ to $0.4$.
The phase space at this resonance is largely regular; there is no discernible
chaotic zone until the eccentricity exceeds $\sim 0.25$.

\item[--]
For $\mean{e}$ exceeding a critical value (which is $\sim0.04$ for the 2:1
Neptune resonance), there are two types of librating orbits:
(i) symmetric librators whose perihelia librate (with large amplitude)
with a mean value $180^\circ$ away from Neptune, and
(ii) asymmetric librators whose perihelia librate (with amplitude smaller
than $\sim 90^\circ$) about a mean value which depends upon $\mean{e}$.
(Surprisingly, the orbit with aphelion exactly at Neptune's longitude is
unstable: it coincides with the separatrix that divides the phase space
into the symmetric and asymmetric libration zones.)
The existence of such asymmetric librations has been noted previously
(\eg \cite{Message58}, \cite{Beauge94}), but the work here provides the first
quantitative analysis for the specific case of Neptune's resonances.

\item[]
Figures 4(e,f) show examples of these two types of orbits.
Note that for the asymmetric librators, there are two independent
centers of libration, offset equally on either side of Neptune's longitude.
Upon inspection of the behavior of the resonance angle $\phi$, it is
found that these centers of perihelion libration (relative to Neptune)
vary from $(\pm)180^\circ$ for $\mean{e}\approx 0$ to $(\pm)50^\circ$
for $\mean{e}\approx0.4$.

\item[--]
The phase space for the symmetric librators shrinks rapidly with increasing
$\mean{e}$, while that for the asymmetric librators increases slightly.

\item[--]
At small values of $\mean{e}$, there is no evidence for a chaotic zone near
this resonance, but for $\mean{e}\gta 0.25$, the largest amplitude librators
become chaotic. A thin chaotic zone also appears in the neighborhood of the
separatrix that separates the asymmetric librations from the symmetric
librations.

\item[--]
The eccentricity variations of the resonance-locked orbits have amplitude
$\delta e\lta0.01$. (Indeed, within the planar circular restricted three-body
model, an initially circular test particle orbit at the 2:1 resonance has its
eccentricity pumped up to only $\sim0.03$.) The eccentricity variations on the
chaotic orbits are generally of considerably larger magnitude.

\item[--]
Figure 4(g) shows libration amplitude vs.~libration period. For the asymmetric
(small amplitude) librators, the libration period decreases monotonically with
$\mean{e}$ and increases with libration amplitude (reminiscent of the standard
non-linear pendulum). The libration period becomes indefinitely large at the
separatrix between the asymmetric and symmetric librators.  The small amplitude
libration periods are several tens-of-thousands of years long.
\end{itemize}

\bigskip
\noindent{\it The 3:1 Neptune resonance}

The synodic period of a test particle near the 3:1 resonance is half its
orbital
period.  Thus, conjunctions with Neptune occur approximately every 248 years.
The properties of the phase space near the 3:1 Neptune resonance (cf.~Figure 3)
are very similar to that near the 2:1 resonance; some specifics are noted
below.

\begin{itemize}

\item[--]
The resonance half-width, $\Delta a$, increases --- from $\lta 0.001$ to
$\sim 0.016$ --- as $\mean{e}$ increases from $\sim0$ to $0.4$.
The phase space at this resonance is regular; there is no discernible
chaotic zone for this range of eccentricities.

\item[--]
For librating orbits, perihelion and aphelion occur at approximately the same
longitude relative to Neptune. As in the 2:1 resonance, there are two types of
librating orbits -- the symmetric and asymmetric librators; however, the
asymmetric librations appear only at $\mean{e}$ exceeding $\sim 0.13$.

\item[--]
The small amplitude libration periods are very long ($\sim 0.4$ Myr) for low
eccentricity $e \lta 0.1$ orbits, but considerably shorter (30,000--80,000 yr)
for higher eccentricities.

\end{itemize}

\section{Summary and Discussion}

Numerical studies of the stability of low eccentricity, low inclination orbits
of small objects in the trans-Neptunian Kuiper Belt subject to the
gravitational perturbations of the giant planets have shown that the inner
edge of the Kuiper Belt is at about 34 AU heliocentric distance; beyond 34 AU
and up to about 42 AU there are interspersed regions of stability and
instability [\cite{LevisonDuncan93}, \cite{HolmanWisdom93}].  This structure
bears a complex correlation with the locations of Neptune orbital resonances.
Other studies [\cite{Malhotra93}, \cite{Malhotra95}] suggest that early in
the history of the Solar system, the majority of Kuiper Belt objects were swept
into eccentric orbits in narrow zones located at Neptune's orbital resonances,
and that the regions in-between the resonances would have been largely cleared
of residual planetesimals.  Recent and ongoing observational surveys of the
outer Solar system indicate the presence of a large population of small bodies
beyond Neptune [\cite{JewittLuu95}]. These are quite likely the source of short
period comets [\cite{Duncanetal88}, \cite{Quinnetal90}]. Furthermore, their
orbital distribution is likely to hold clues to the formation and early
dynamical evolution of the outer Solar system [\cite{Malhotra95}].  All of
these considerations have motivated the present study.

Using the planar circular restricted three body model (with the Sun, Neptune
and a test particle), I have described in this paper the basic phase space
structure in the neighborhood of Neptune's exterior orbital resonances.  The
details may be gleaned from Figures 3--10. A succinct summary of these results
is given in Figure 11 which shows the locations and widths of the stable
resonance libration zones. In general, these stable zones are bounded by
chaotic layers of thickness that generally increases with eccentricity and
decreases with distance from Neptune. (As the mean semimajor axis and
eccentricity of chaotic orbits is not well defined, it is not possible to
represent the chaotic layers in such a figure.)

The planar circular restricted three body model is the simplest dynamical model
for the orbital dynamics of small objects in the Kuiper Belt. Although this
model may appear oversimplified, it provides a reasonable description and
explanation for much of the dynamical behaviors found in the numerical studies
mentioned above.\footnote{Those
numerical studies assumed a realistic three-dimensional physical model in
which test particles were perturbed by all the four giant planets, Jupiter,
Saturn, Uranus and Neptune; the planets' orbits were integrated
self-consistently under their mutual gravitational perturbations.}
For example, the location of the inner edge of this Belt at $\sim$(33--34) AU
is
readily understood in terms of orbital instability induced by overlapping first
order Neptune resonances; the additional perturbations due to Neptune's
non-circular, inclined, and time-varying orbit and the perturbations of the
other planets do not significantly change the location of this boundary.

The relatively isolated first and second order Neptune resonances beyond 34 AU
(in semimajor axis) provide narrow stable libration regions for the long term
storage of Kuiper Belt objects in eccentric (often Neptune-crossing) orbits.
This paper provides first approximation estimates for the locations and widths
of these regions, and for the dynamical properties of resonant orbits.
It also provides a direct visualization --- in two-dimensional
surfaces-of-section -- of the global phase space structure
(\ie the chaotic and stable regions) in the vicinity of orbital resonances
A characteristic of stable resonant orbits is the libration of the perihelion
about a longitude well removed from Neptune's location. Typical libration
periods are several tens-of-thousands of years.  The libration zones are
generally surrounded by narrow chaotic zones where orbits are unstable on
timescales of a few libration periods, or $\sim 10^5$ yr. At the boundary
between the stable resonance libration zone and the chaotic zone, the timescale
for orbital instability may be exceedingly long.
This paper also provides a quantitative analysis of libration periods
and their dependence on libration amplitudes and orbital eccentricities
at all the major exterior mean motion resonances of Neptune.

Of course, the planar circular restricted three body model does not provide
a complete picture.
Indeed, I expect that it underestimates the extent of the chaotic zones and
overestimates the sizes of the stable libration zones near Neptune resonances.
But perhaps the most important missing element is the effect of secular
resonances [see \eg \cite{Knezevicetal91}] on the long term dynamics of Kuiper
Belt objects.  For example, in some regions, secular resonances produce
significant inclination excitation; in such cases, the assumption of planarity
becomes a poor approximation. [Pluto's orbit is a case in point
(\cite{NacozyDiehl78}).]
In order to discern these effects, one has to build much more elaborate
analytical models or resort to extensive numerical integrations.

As this work was being completed, I became aware of two preprints
[\cite{Morbidellietal95} and \cite{Duncanetal95}] on the same subject.
These use different approaches to the problem: the present work
has focussed on the phase space near mean motion resonances with Neptune only
and gives results that are exact but for a highly simplified physical model;
\cite{Morbidellietal95} discuss the dynamics near mean motion resonances as
well as secular resonances using primarily approximate semi-analytic models;
and \cite{Duncanetal95} discuss the dynamics near mean motion resonances,
secular resonances as well as non-resonant regions using a purely numerical
approach.
A comparison of the present work with these two preprints follows.

Morbidelli \etal (1995) describe semi-analytic and numerical
investigations that include the effects of all four giant planets on the
dynamics of test particles in the Kuiper Belt.  A comparison of their
results for the locations and widths of mean motion resonances (see
their Figure 1) with the estimates obtained here (cf.~Figure 11) shows
that their models significantly overestimate the sizes of the resonance
libration regions.  Morbidelli {\it et al.}'s
estimates of the resonance widths were obtained by averaging the planar,
circular restricted problem of the four giant planets plus a test
particle over fast variables, retaining only the dependence on the
resonance angle.  Thus, their analysis does not account for the
overlapping secondary resonances at the edges of the mean motion
resonance libration zones that produce a chaotic layer and cause the
libration regions to shrink rapidly with increasing eccentricity.
This is the explanation for the differences between their figure 1 and
figure 11 here.  In addition,
Morbidelli \etal find a singularity in the averaged Hamiltonian that
causes a great increase in the resonance width near Neptune-crossing
values of the eccentricity.  However, this is an artifact, and does not
appear in the unaveraged problem, as is obvious from the
surfaces-of-section shown in the present work.  The technique I used here
does not make any averaging approximations, but uses the full unaveraged
Hamiltonian for the circular planar restricted problem, albeit without
the secular effects of Jupiter, Saturn and Uranus.  Morbidelli {\it et al.}'s
models also show that secular effects are negligible near mean motion
resonances except in the case of the 3:2 Neptune resonance.  In the
special case of the 3:2 resonance, as mentioned above, \cite{Knezevicetal91}
had previously shown the existence of the  $\nu_{18}$ secular resonance
embedded within the mean motion resonance; in addition, analysis of
Pluto's orbit had also previously revealed the existence of another
resonance characterized by the libration of the argument of perihelion
(\cite{NacozyDiehl78}).
Both of these resonances have the effect of exciting the inclination of
the test particle. These resonances are not modeled in the present work,
but Morbidelli \etal describe in some detail the location and widths of
both these and other secular resonances.
However, note that because most regions near secular resonances are regions of
orbital instability, they are not likely to be of importance as reservoirs
of Kuiper Belt objects although they may drive their transport to small
heliocentric distances at the present epoch.

Duncan \etal (1995) have performed quite extensive numerical integrations of
test particle orbits subject to the gravitational perturbations of the
four giant planets.
They describe their results in terms of the stability (over their
integration time of 1--4 Byr) of orbits with particular initial
semimajor axes, eccentricities and inclinations. Their main conclusions
are consistent with the results of the present work: namely, that
particles with low initial inclination ($\sim1^\circ$) and initial
perihelion distance less than $\sim35$ AU are unstable on timescales
short compared to the age of the Solar system, except that
particles librating in low order mean motion resonances with Neptune
remain phase-protected from close encounters with that planet.
They have also analyzed the 3:2 Neptune resonance in more detail for the
particular case with initial eccentricities of 0.2 and found three regimes:
stable orbits deep in resonance with small libration amplitudes,
$\lta70^\circ$;
an intermediate region with libration amplitudes in the range
$70^\circ-130^\circ$ where
the timescale for instability is $\sim10^9$ yr;  and highly unstable
orbits for libration amplitudes exceeding $\sim130^\circ$ with stability
timescales of $\lta10^8$ yr. Similar dynamics is found for all the
Neptune resonances studied in this paper.
In comparison, the present work has shown that the region of stability deep in
resonance as well as the highly chaotic region at large amplitudes  are both
explained by the perturbations of Neptune alone (assumed to be
on a circular orbit); the intermediate regime where the timescale for
instability is $\sim10^9$ yr is very small in the restricted three body model
but evidently expands greatly (at the expense of the stable libration zone)
in the full N-body model for the outer Solar system. [In this context, we note
that if Malhotra's (1995) theory for the orbital distribution of Kuiper Belt
objects is correct, then this intermediate libration amplitude regime
which allows a long-term leakage of resonant objects may be the primary source
of the Jupiter-family short-period comets.] Duncan \etal also find
that the resonance-protection fails only at inclinations exceeding
$\sim25^\circ$. Thus, we conclude that the planar circular restricted three
body model used in the present work provides a fairly good description of
these main trends. A clear advantage of this simple model is that its analysis
is quite ``inexpensive'' compared to the months of CPU time expended in the
numerical simulations by Duncan \etal
Furthermore, the numerical
integration of orbits for the age of the Solar system at sufficiently
high resolution in the space of initial conditions remains beyond the
reach of present-day computers, as acknowledged by Duncan \etal
Thus, the global picture of the phase space structure obtained
in the present work provides an inexpensive yet reasonably good baseline for
understanding the dynamics in the trans-Neptunian Solar system,
and is particularly useful near mean motion resonances.

\bigskip

This research was done while the author was a Staff Scientist at the
Lunar and Planetary Institute which is operated by the Universities Space
Research Association under contract no.~NASW-4574 with the National Aeronautics
and Space Administration. Partial support for this work was also provided by
NASA's Origins of Solar Systems Research Program under grant no.~NAGW-4474.
This paper is Lunar and Planetary Institute Contribution no.~870.

%%%%%%%%%%%%%%%%%%%%%%%%%%%%%%%%%%%%%%%%%%%%%%%%%%%%%%%%%%%%%%%%%%%%%%%%%%%%%%%

\clearpage

\centerline{FIGURE CAPTIONS}
\bigskip

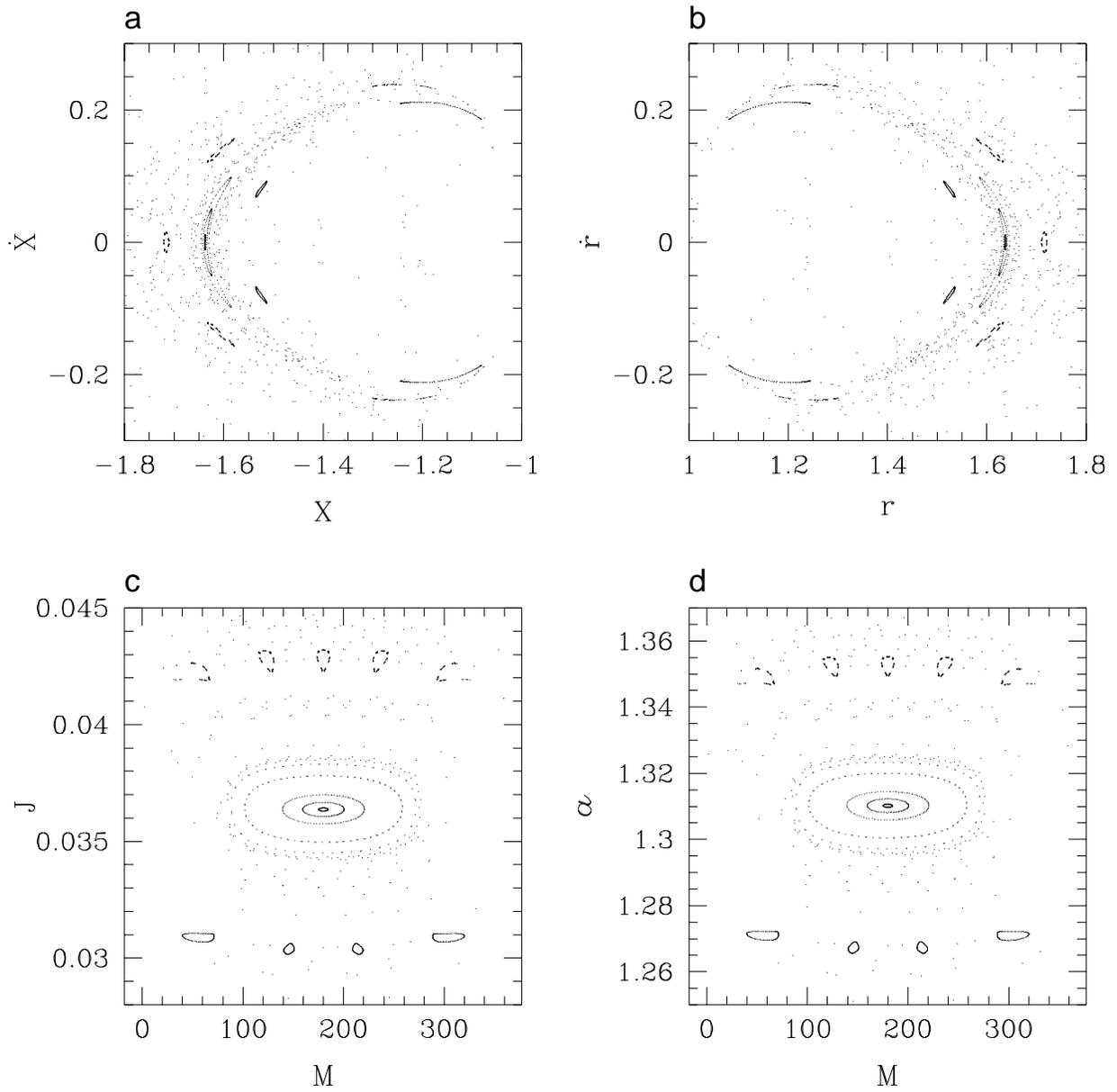
\noindent{Fig.~1---}
The phase space structure in the neighborhood of the Neptune-Pluto 3:2
orbital resonance determined from the circular planar restricted three-body
model. The same surface-of-section is shown in different variables:
(a) cartesian variables, $(X,\dot X)$; (b) plane polar variables, $(r,\dot r)$;
(c) Delaunay variables (mean anomaly $M$, canonical momentum
$J=\sqrt{a}(1-\sqrt{1-e^2})$). (d) is a pseudo-surface-of-section since the
semimajor axis, $a$, is not a canonical variable. However, qualitatively it
is very similar to the surface-of-section (c), and it provides a visualization
of the resonance width in terms of the semimajor axis and the amplitude of the
perihelion libration.
\bigskip
%%%%%%%%%%%%%%%%%%%%%%%%%%%%%%%%%%%%%%%%%%%%%%%%%%%%%%%%%%%%%%

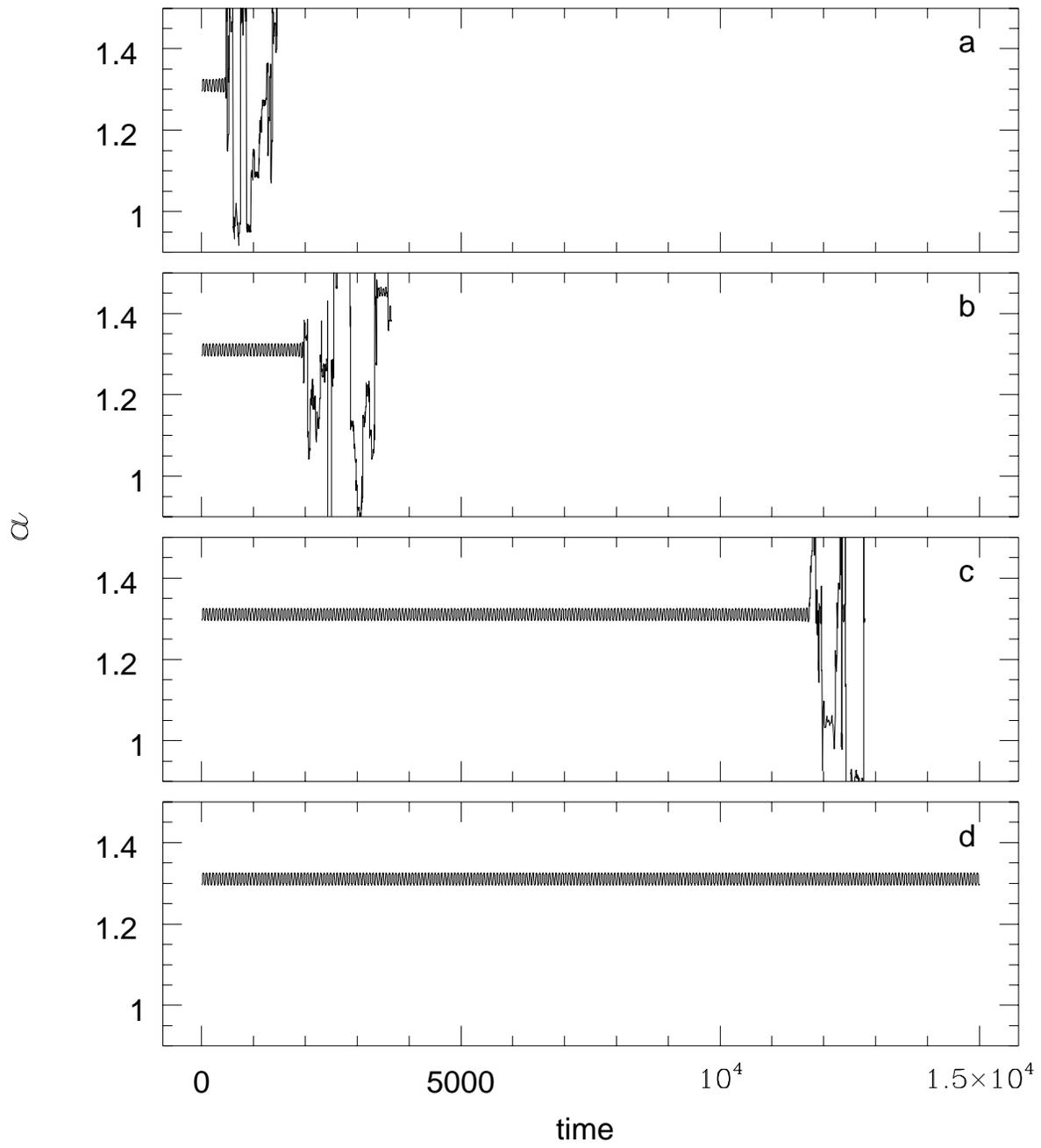
\noindent{Fig.~2---}
The time variation of the semimajor axis for orbits with $e\approx0.25$ near
the boundary between the stable libration zone and the chaotic zone at the 3:2
Neptune resonance (cf.~Figure 1).
(The unit of length is Neptune's semimajor axis, and the unit of time is
Neptune's orbital period, approximately 165 yr.)
The Jacobi integral for all four orbits is $C=2.9798$;
the initial conditions $(X,\dot X,Y,\dot Y)$ for the four orbits are
(a)---(-1.6100,0.0,0.0,0.92442),
(b)---(-1.6105,0.0,0.0,0.92508),
(c)---(-1.6106,0.0,0.0,0.92521),
(d)---(-1.6110,0.0,0.0,0.92574).
These orbits are all chaotic and are shown in order of decreasing distance to
the stability boundary.
The orbits in (a), (b) and (c) become unstable after 7, 31 and $\sim 200$
librations (\ie $\sim8\times10^4$, $\sim3\times10^5$ and $\sim2\times10^6$ yr),
respectively.
\bigskip
%%%%%%%%%%%%%%%%%%%%%%%%%%%%%%%%%%%%%%%%%%%%%%%%%%%%%%%%%%%%%%

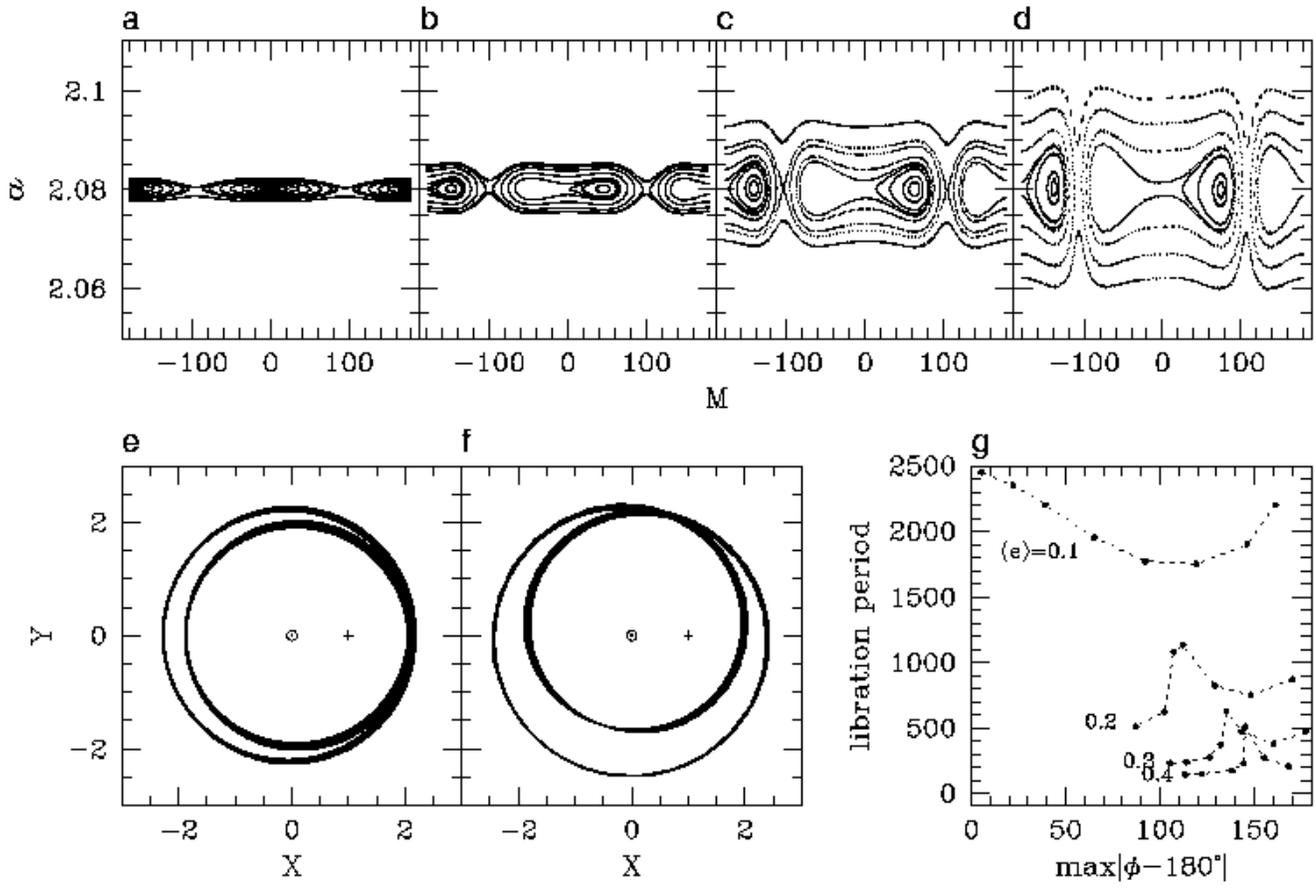
\noindent{Fig.~3---}
The 3:1 Neptune resonance located at $a_{res}=2.08008$ (\ie 62.6 AU).
The phase space for orbits with (a) $C=3.3508, \mean{e}=0.1$;
(b) $ C=3.3070,\mean{e}=0.2$; (c) $C=3.2324,\mean{e}=0.3$;
(d) $ C=3.1244,\mean{e}=0.4$. The phase space is largely regular at this
resonance. The width of the libration zone increases in with $\mean{e}$.  For
$\mean{e}\gta0.13$, the libration zone splits into two zones of asymmetric
librations surrounded by a narrow zone of large-amplitude symmetric librations.
Panel (e) shows a symmetrically librating orbit (with $e\approx0.1$) in the
rotating frame; panel (f) shows an asymmetrically librating orbit (with
$e\approx0.2$). The symbols $\odot$ and $+$ indicate the fixed locations of
the Sun and Neptune in the rotating frame.
Panel (g) shows the libration period (in units of Neptune's orbital period) as
a function of the maximum excursion of the resonance angle $\phi$ from
$180^\circ$.  Note that the center of the asymmetric librators moves away from
``$\phi=180^\circ$'' with increasing $\mean{e}$, as indicated by the beginning
of the curves labeled by the different values of $\mean{e}$. The libration
period increases with libration amplitude for the asymmetric librators, but
decreases with amplitude for the symmetric (large amplitude) librators; at the
separatrix on the boundary between the symmetric and asymmetric librations, the
libration period becomes indefinitely large (although this is not quite obvious
due to the finite resolution in this figure).
\bigskip
%%%%%%%%%%%%%%%%%%%%%%%%%%%%%%%%%%%%%%%%%%%%%%%%%%%%%%%%%%%%%%

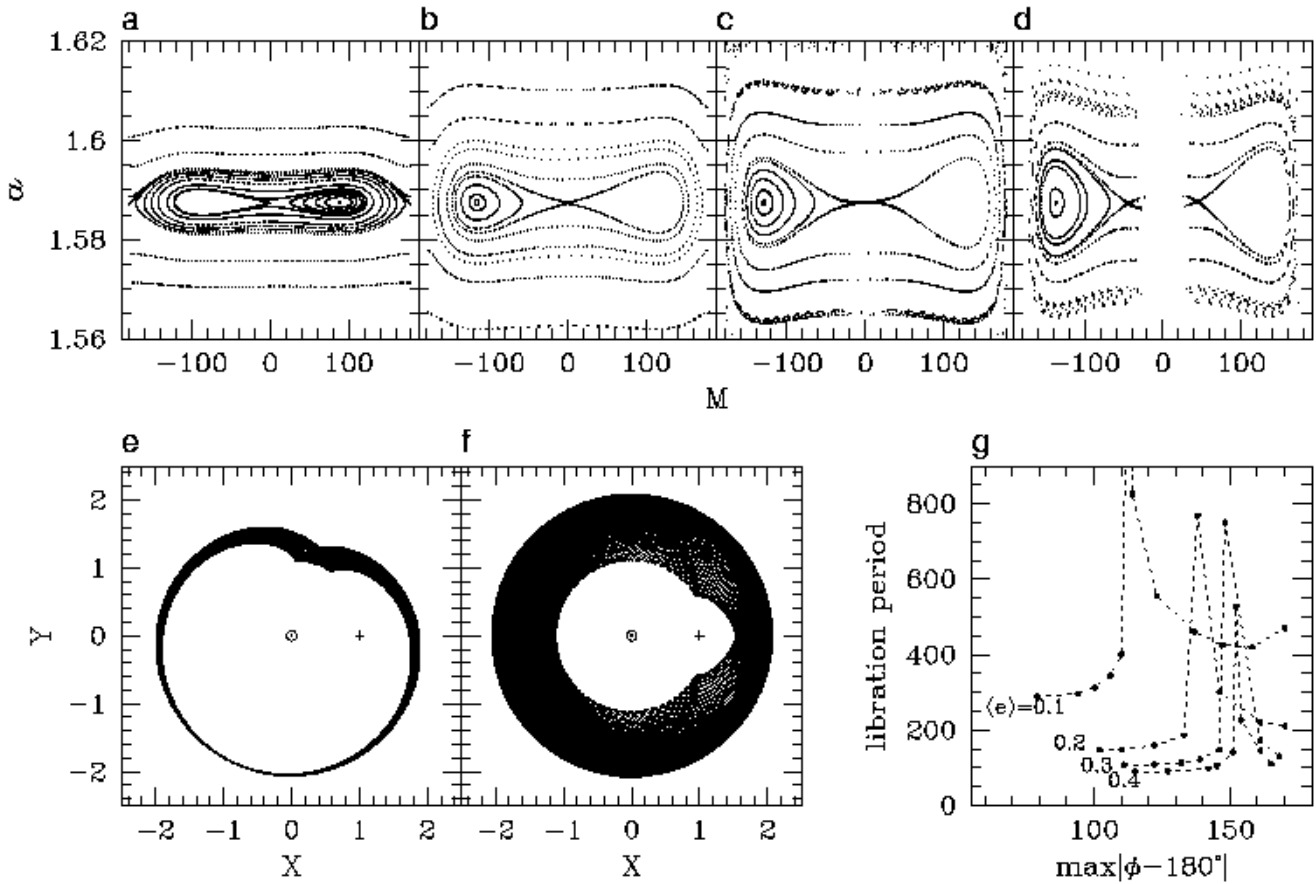
\noindent{Fig.~4---}
The 2:1 Neptune resonance located at $a_{res}=1.5873$ (\ie 47.8 AU).
The phase space for orbits with (a) $C=3.1272, \mean{e}=0.1$;
(b) $ C=3.0989,\mean{e}=0.2$; (c) $C=3.0338,\mean{e}=0.3$;
(d) $ C=2.9395,\mean{e}=0.4$. The phase space is largely regular at this
resonance; a chaotic zone appears at large amplitude librations for
eccentricity
exceeding $\sim 0.25$. The width of the libration zone increases in with
$\mean{e}$.  For $\mean{e}\gta0.04$, the libration zone splits into two zones
of
asymmetric librations surrounded by a narrow zone of large-amplitude symmetric
librations.  Panels (e) and (f) show a small-amplitude asymmetrically librating
orbit and a large amplitude symmetrically librating orbit (both with
$e\approx0.3$) in the rotating frame. (The motion of the test particle over a
complete libration period is traced in these figures.) Note that in both types
of orbits the perihelion occurs away from Neptune's longitude.
Panel (g) shows the libration period (in units of Neptune's orbital period) as
a function of the maximum excursion of the resonance angle $\phi$ from
$180^\circ$.  The center of the asymmetric librators is indicated by the
beginning of the curves labeled by the different values of $\mean{e}$. The
libration period increases with libration amplitude for the asymmetric
librators, but decreases with amplitude for the symmetric (large amplitude)
librators; at the separatrix on the boundary between the symmetric and
asymmetric librations, the libration period becomes indefinitely large.
\bigskip
%%%%%%%%%%%%%%%%%%%%%%%%%%%%%%%%%%%%%%%%%%%%%%%%%%%%%%%%%%%%%%

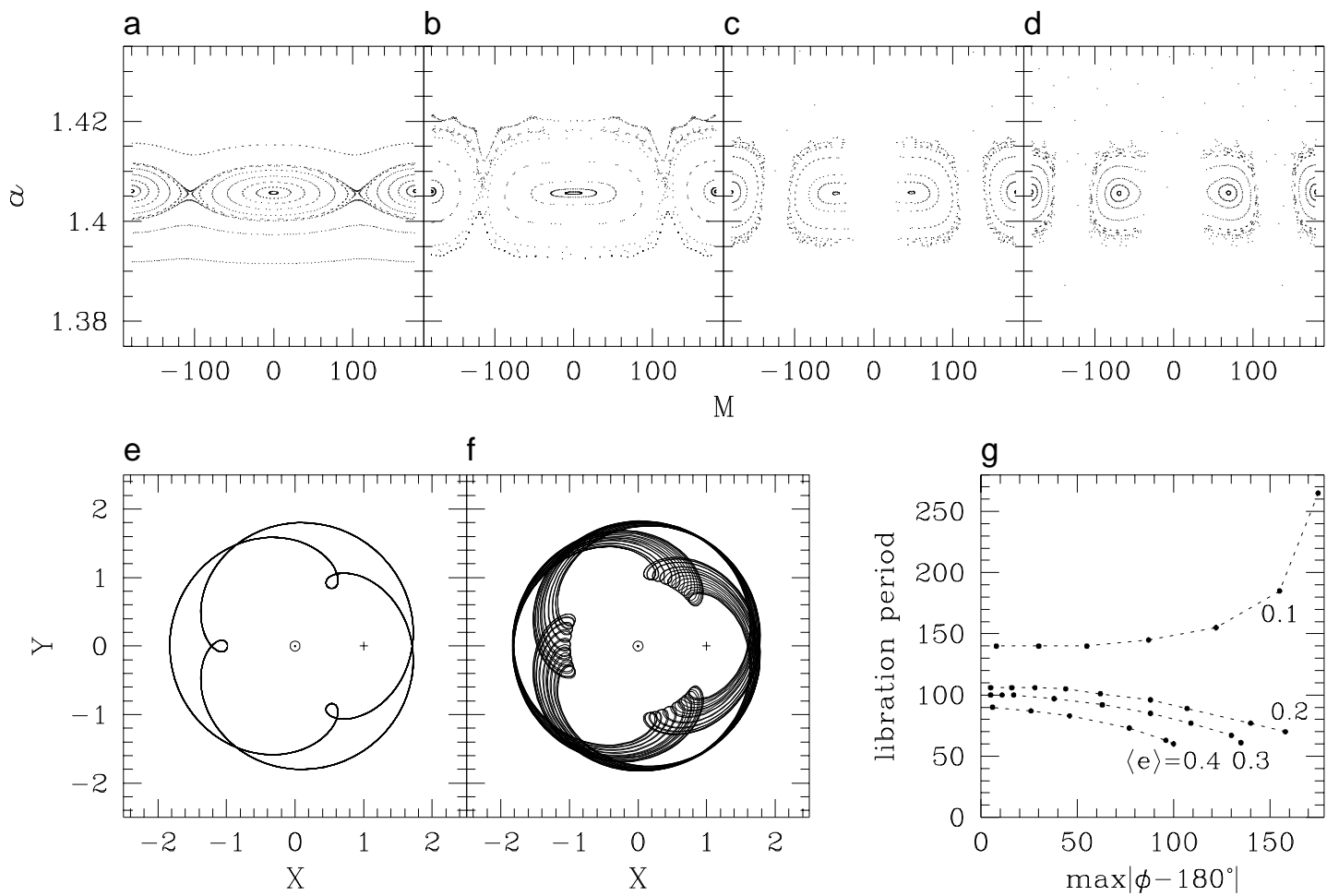
\noindent{Fig.~5---}
The 5:3 Neptune resonance located at $a_{res}=1.4057$ (\ie 42.3 AU).
The phase space for orbits with (a) $C=3.0709, \mean{e}=0.1$;
(b) $ C=3.0349,\mean{e}=0.2$; (c) $C=2.9739,\mean{e}=0.3$;
(d) $ C=2.8848,\mean{e}=0.4$.
The width of the libration zone increases for $\mean{e}$ up to $\sim 0.2$, but
then shrinks with increasing $\mean{e}$ as the larger amplitude resonant orbits
dissolve into the chaotic zone.
Note that for $\mean{e}\gta0.3$, the formerly period-2 exact resonant orbit in
the surface-of-section becomes a period-3 orbit;
this is merely a reflection of the fact that at large, Neptune-crossing
eccentricities the perihelion loop results in two conjunctions with Neptune in
rapid succession [cf.~panel (e) and (f)].
Panels (e) and (f) show the periodic orbit at exact resonance, and an orbit
librating with finite amplitude, respectively (both with $e\approx0.3$) in the
rotating frame. (The motion of the test particle over a complete libration
period is traced in these figures.) Note that in both types of orbits the
perihelion occurs away from Neptune's longitude.
Panel (g) shows the libration period (in units of Neptune's orbital period) as
a function of the libration amplitude of the resonance angle $\phi$.
The libration period increases with libration amplitude only for small
$\mean{e}$; it decreases with amplitude for larger $\mean{e}$.
\bigskip
%%%%%%%%%%%%%%%%%%%%%%%%%%%%%%%%%%%%%%%%%%%%%%%%%%%%%%%%%%%%%%

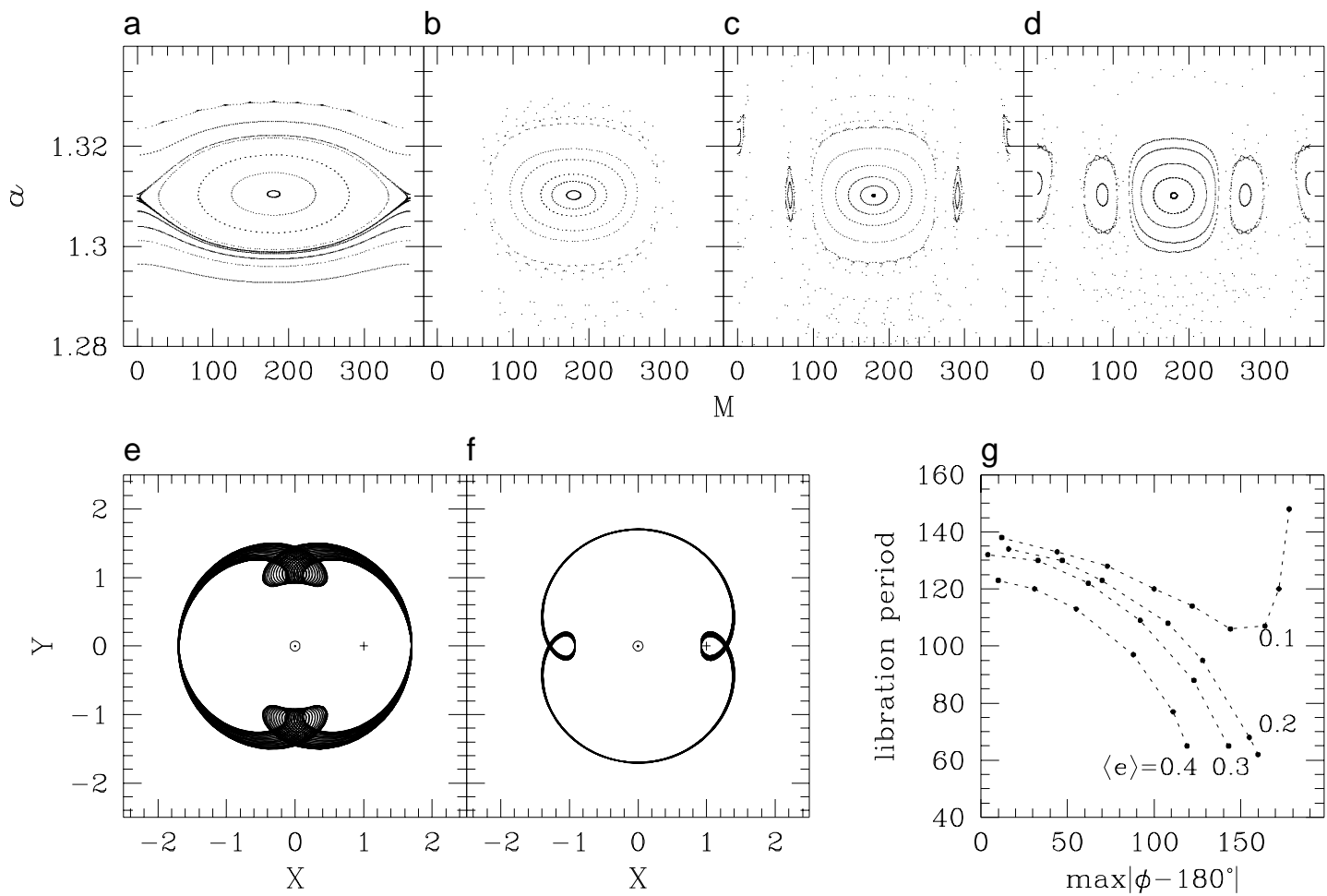
\noindent{Fig.~6---}
The 3:2 Neptune resonance located at $a_{res}=1.3104$ (\ie 39.4 AU).
The phase space for orbits with (a) $C=3.0411, \mean{e}=0.1$;
(b) $ C=3.0065,\mean{e}=0.2$; (c) $C=2.9470,\mean{e}=0.3$;
(d) $ C=2.8616,\mean{e}=0.4$. The width of the main libration zone decreases
with increasing $\mean{e}$, as the chaotic zone surrounding the libration
region
expands.  Panel (e) shows an example of a stably librating orbit (with
$e\approx0.3$) in the rotating frame; note that this is an
``aphelion librator'': the test particle is near aphelion when passing
Neptune's
longitude, and its perihelion librates about $\pm90^\circ$ away from Neptune.
For $\mean{e}\gta0.3$, a new libration zone appears in which the librating
orbits have a perihelion near Neptune.  This is indicated by the appearance of
a
chain of two libration islands (distinct from the main resonance zone) in
panels
(c)--(d); these are the ``perihelion librators'' [cf.~panel(f)]; The new
libration zone increases with increasing $\mean{e}$.  An example of a
perihelion
librator is shown in panel (f); note that one of the two perihelion loops in
each synodic period encompasses Neptune.
Panel (g) shows the libration period (in units of Neptune's orbital period) of
the aphelion librators as a function of the libration amplitude of the
resonance
angle $\phi$.  The libration period generally decreases with libration
amplitude, except for very small eccentricity ($e\lta0.1$), large-amplitude
librators.
\bigskip
%%%%%%%%%%%%%%%%%%%%%%%%%%%%%%%%%%%%%%%%%%%%%%%%%%%%%%%%%%%%%%

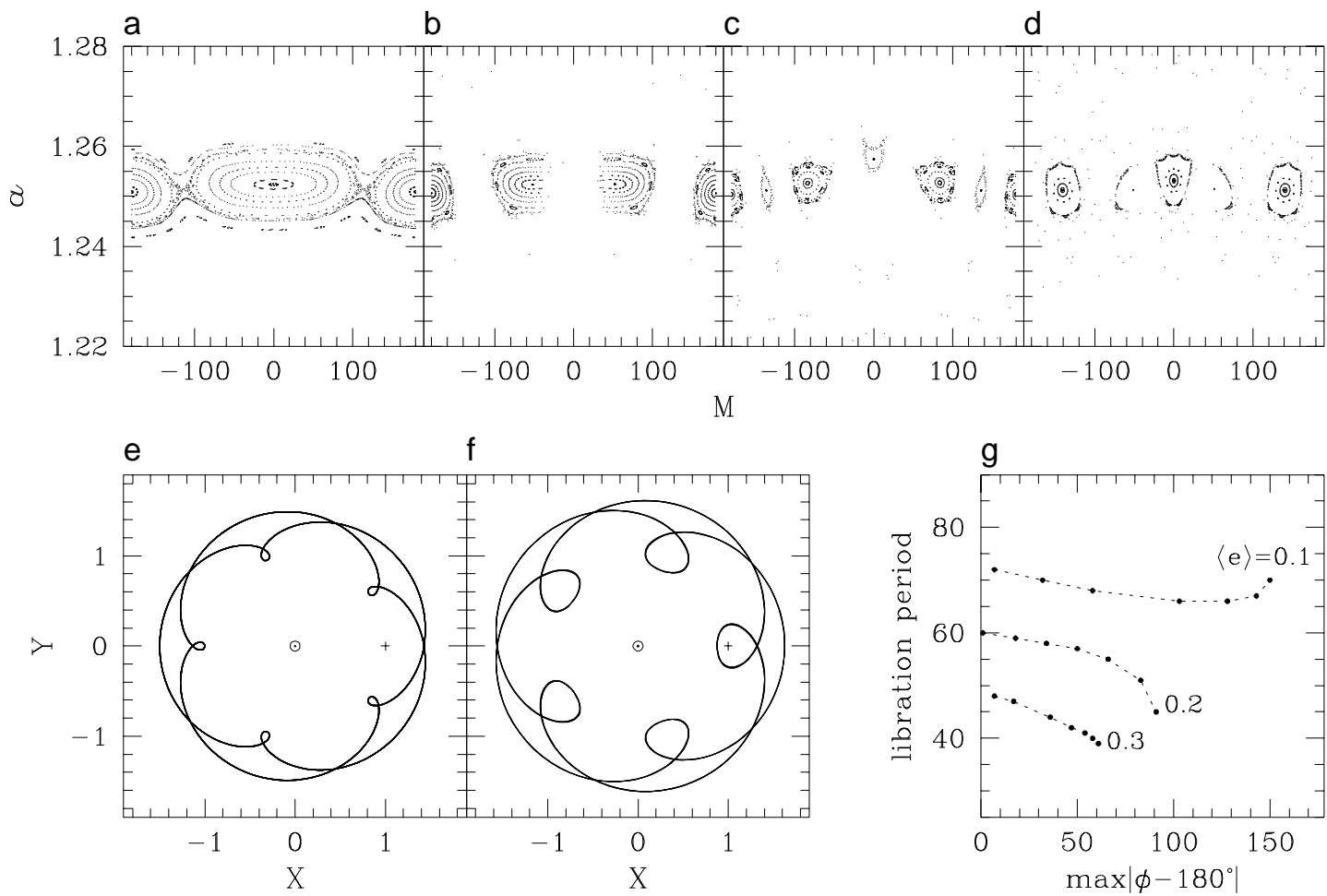
\noindent{Fig.~7---}
The 7:5 Neptune resonance located at $a_{res}=1.2515$ (\ie 37.7 AU).
The phase space for orbits with (a) $C=3.0255, \mean{e}=0.1$;
(b) $ C=2.9913,\mean{e}=0.2$; (c) $C=2.9334,\mean{e}=0.3$;
(d) $ C=2.8496,\mean{e}=0.4$. The width of the main libration zone (for
aphelion
librators) decreases rapidly with increasing $\mean{e}$, as the chaotic zone
surrounding the libration region expands. In fact, for $\mean{e}\ge 0.3$
[cf.~panel (d)] the aphelion libration zone has completely disappeared.
A libration zone for perihelion librators appears at $e\approx0.3$ and expands
slightly with increasing $\mean{e}$.
Panels (e) and (f) show examples of an aphelion librator with $e\approx0.2$,
and a perihelion librator with $e\approx0.3$, respectively.
Panel (g) shows the libration period (in units of Neptune's orbital period) of
aphelion librators as a function of the libration amplitude of the resonance
angle $\phi$.  The libration period generally decreases with libration
amplitude, except for very small eccentricity ($e\lta0.1$), large-ampitude
librators.
\bigskip
%%%%%%%%%%%%%%%%%%%%%%%%%%%%%%%%%%%%%%%%%%%%%%%%%%%%%%%%%%%%%%

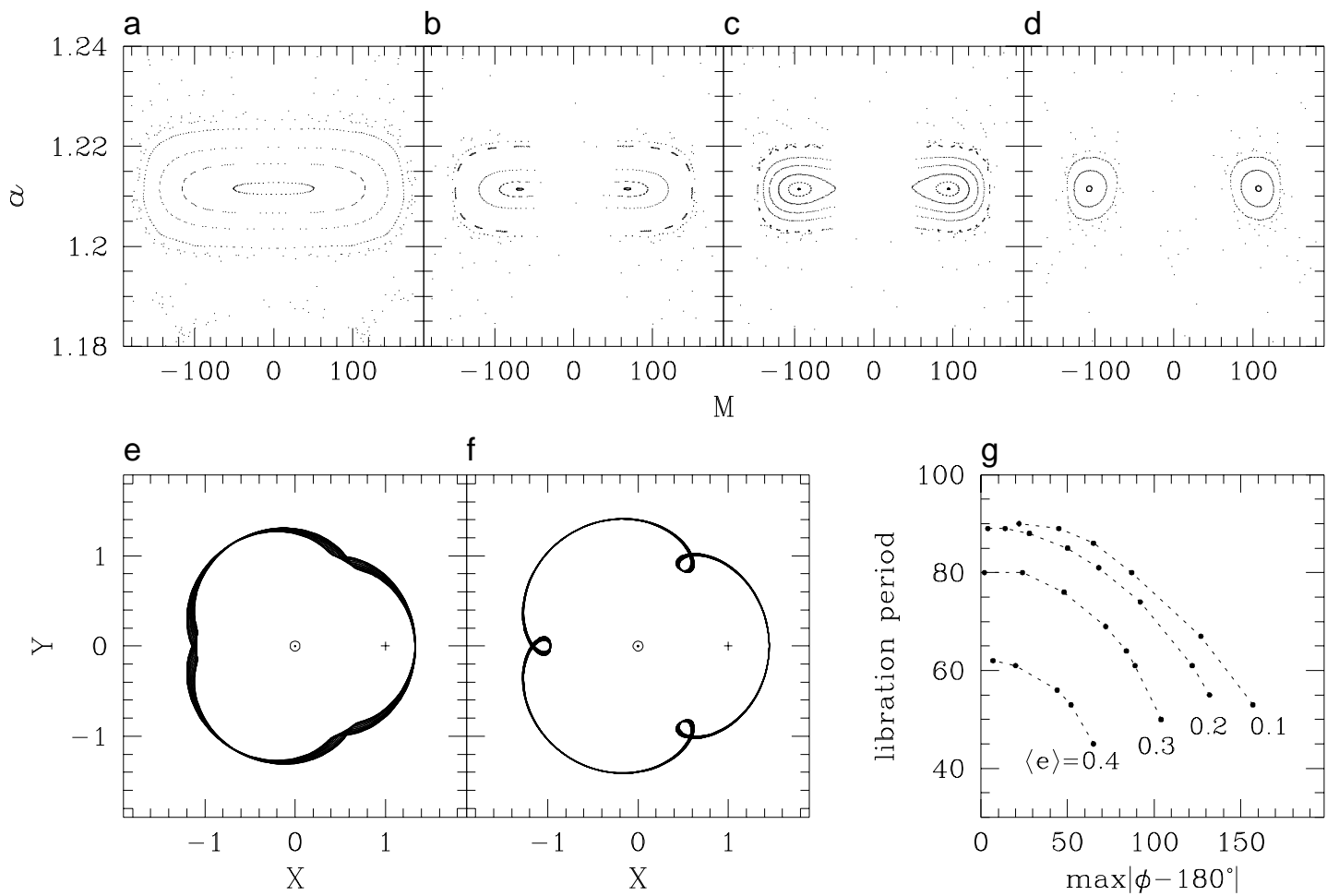
\noindent{Fig.~8---}
The neighborhood of the Neptune 4:3 orbital resonance located at
$a_{res}=1.2114$ (\ie 36.5 AU).
The phase space for orbits with (a) $C=3.0158, \mean{e}=0.1$;
(b) $ C=2.9824,\mean{e}=0.2$; (c) $C=2.9253,\mean{e}=0.3$;
(d) $ C=2.8430,\mean{e}=0.4$. The libration zone shrinks rapidly with
$\mean{e}$ as the larger amplitude librators dissolve into the chaotic zone.
For $\mean{e}>0.175$, the formerly period-1 orbit corresponding to exact 4:3
resonance bifurcates into a period-2 orbit in the surface-of-section;
this is merely a reflection of the fact that for a Neptune-crossing value
of the eccentricity, the perihelion loop causes two conjunctions with
Neptune in each synodic period.
Panel (e) and (f) show examples of small amplitude librating orbit with
$e\approx0.1$ and $e\approx0.2$, respectively.
Panel (g) shows the period of libration as a function of the libration
amplitude of the resonance angle, $\phi$ (for $\mean{e}=0.1,0.2,0.3$ and
$0.4$). The period as well as the maximum amplitude for stable librations
decreases with increasing $\mean{e}$.
\bigskip
%%%%%%%%%%%%%%%%%%%%%%%%%%%%%%%%%%%%%%%%%%%%%%%%%%%%%%%%%%%%%%

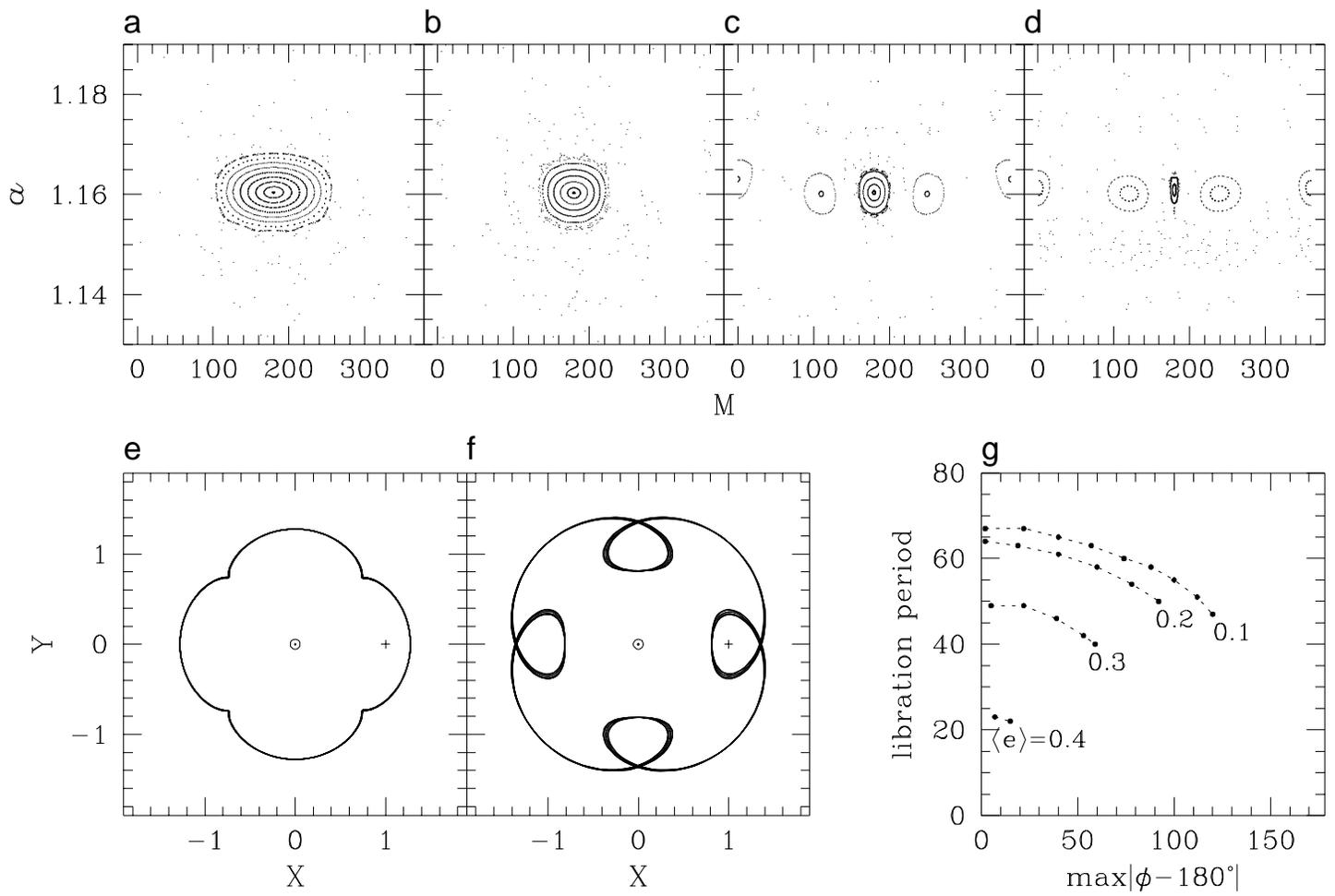
\noindent{Fig.~9---}
The neighborhood of the Neptune 5:4 orbital resonance located at
$a_{res}=1.1604$.
The phase space for orbits with (a) $C=3.0054, \mean{e}=0.1$;
(b) $ C=2.9727,\mean{e}=0.2$; (c) $C=2.9170,\mean{e}=0.3$;
(d) $ C=2.8364,\mean{e}=0.4$. Notice that the main libration zone (for aphelion
librators) shrinks rapidly with $\mean{e}$.  A libration zone for perihelion
librators appears for $\mean{e}>0.2$, as indicated by the appearance of a chain
of two libration islands in (c)--(d); this zone increases with increasing
$\mean{e}$.  Panel (e) shows the stable periodic orbit with $e\approx0.3$ at
the
5:4 resonance (an aphelion librator) in the rotating frame coordinates; panel
(f) shows an example of a perihelion librator with $e\approx0.3$.  Notice the
four-fold symmetry in the rotating frame.  Panel (g) shows the period of
libration for the aphelion librators as a function of the libration amplitude
of
the resonance angle (for $\mean{e}=0.1,0.2,0.3$ and $0.4$)
\bigskip
%%%%%%%%%%%%%%%%%%%%%%%%%%%%%%%%%%%%%%%%%%%%%%%%%%%%%%%%%%%%%%

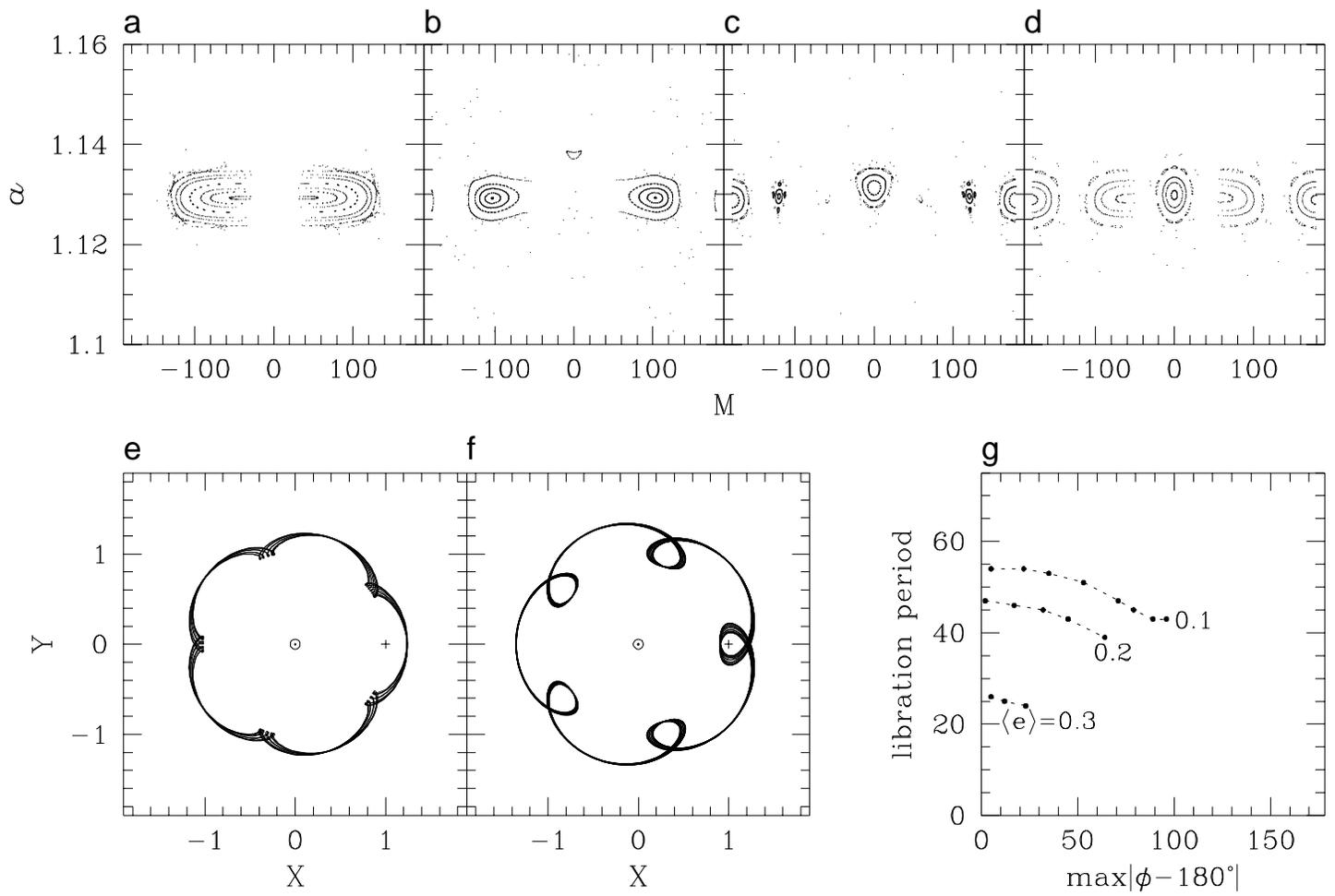
\noindent{Fig.~10---}
The neighborhood of the Neptune 6:5 orbital resonance located at
$a_{res}=1.1292$ (\ie 34.0 AU).
The phase space for orbits with (a) $C=3.0002, \mean{e}=0.1$;
(b) $ C=2.9680,\mean{e}=0.2$; (c) $C=2.9131,\mean{e}=0.3$;
(d) $ C=2.8334,\mean{e}=0.4$. Notice that the main libration zone (for aphelion
librators) shrinks rapidly with $\mean{e}$; for $\mean{e}>0.1$ the center of
this libration zone bifurcates into a period-2 orbit; this is simply a
reflection of the fact that the perihelion loop causes two conjunctions with
Neptune in rapid succession during each synodic period.
Also for $\mean{e}>0.1$,
a new libration zone appears for perihelion librators [indicated by the
appearance of a chain of two libration islands with centers at $M=0,\pi$ in
(b)--(d)] whose extent increases with increasing $\mean{e}$.  The increased
complexity of this libration zone in (d) is a reflection of the fact that the
perihelion loops for large amplitude librations are so large that there are
four
conjunctions with Neptune during each synodic period. Panels (e) and (f) show
examples an aphelion librator and of a perihelion librator with $e\approx0.3$
are the 6:5 resonance.  Notice the five-fold symmetry in the rotating frame.
Panel (g) shows the period of libration as a function of the libration
amplitude
of the resonance angle, $\phi$ (for $\mean{e}=0.1,0.2,0.3$ and $0.4$) for the
aphelion librators.
\bigskip
%%%%%%%%%%%%%%%%%%%%%%%%%%%%%%%%%%%%%%%%%%%%%%%%%%%%%%%%%%%%%%

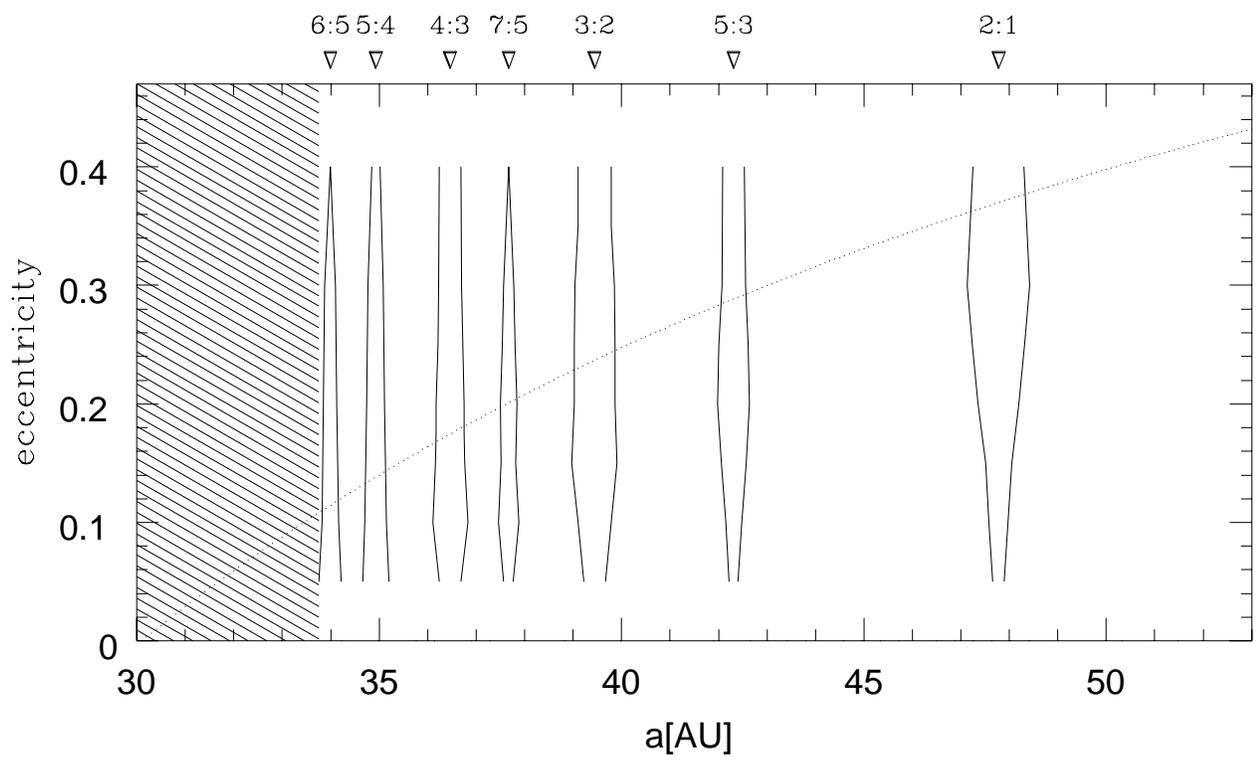
\noindent{Fig.~11---}
The locations and widths of first and second order Neptune resonances in the
Kuiper Belt, as determined by the planar, circular restricted three-body model.
The shaded region on the extreme left indicates the chaotic zone of
``resonance overlap'' in the vicinity of Neptune's orbit. In the region above
dotted line, orbits are Neptune-crossing; these orbits are dynamically
short-lived (due to close encounters with Neptune), unless protected by
orbital resonances.
%%%%%%%%%%%%%%%%%%%%%%%%%%%%%%%%%%%%%%%%%%%%%%%%%%%%%%%%%%%%%%

\end{document}